\documentclass[%
superscriptaddress,
preprint,
amsmath,amssymb,
aps,
]{revtex4-2}

\usepackage{graphicx}
\usepackage{dcolumn}
\usepackage{bm}

\begin{document}
	
	
	\title{Suppression of electroconvection due to van der Waals attraction of polymer additives towards the metal electrode}
	
	\author{Ankush Mukherjee}
	\affiliation{%
		Sibley School of Mechanical and Aerospace Engineering, Cornell University, Ithaca, New York 14853, USA
	}%
	

	\author{Lynden A. Archer}
	\author{Donald L. Koch}%
	\email{dlk15@cornell.edu}
	\affiliation{
		Robert Frederick Smith School of Chemical and Biomolecular Engineering, Cornell University,
		Ithaca, New York 14853, USA
	}%
	
	
	\date{\today}
	
	\begin{abstract}
			 Electroconvection in rechargeable batteries enhances the growth of dendrites at the electrode surface. The addition of low molecular weight polymers to the electrolyte in batteries results in the formation of a thin layer of higher polymer concentration near the electrode. This is due to van der Waals forces of attraction between the metal electrode and the polymers dissolved in the electrolyte. The van der Waals forces act as a restoring body force on the electrolyte and oppose the growth of perturbations. Using linear stability analysis, we show that this  force opposes  electroconvective flow. This increases the critical voltage required for the onset of electroconvection.
	\end{abstract}
	
	\maketitle
	
	
	\section{\label{sec:1}Introduction}

Electroconvection has many applications such as  electrodialysis \cite{rubinstein1997electric,strathmann2010electrodialysis}, desalination \cite{gonzalez2020exploiting,nikonenko2014desalination} and  batteries with liquid electrolytes \cite{fleury1992theory,wang1994formation,fleury1993coupling,huth1995role}. Non-planar electrodeposition in rechargeable lithium batteries  has been studied extensively \cite{tarascon2011issues,xu2004nonaqueous,cheng2017toward,alias2015advances,zheng2020regulating} as it reduces the reversibility of the battery and results in the growth of dendrites, which can grow to the opposite electrode and short the battery \cite{wu2017interplay,rosso2006dendrite}.  Polymers dissolved in the electrolyte are attracted towards the metal electrode due to weak van der Waals forces of attraction between the metal and dissolved polymers. These polymers remain dissolved in the electrolyte and form a thin layer adjacent to the ion-selective surface. This polymer layer near the electrode has a stabilizing effect on electroconvection due to the van der Waals forces which act as a restoring body force that opposes the flow of the electrolyte. To suppress electroconvection, various techniques have been used to modify the electrolyte, such as the use of solid electrolytes and the addition of high molecular weight polymers to increase the viscosity. An advantage of the method used in this study is that low molecular weight polymers can be used, so the properties of the bulk electrolyte remain almost the same, hence ion transport in the bulk electrolyte is not significantly affected. 

At low voltages, ion concentration profiles in batteries are one-dimensional. The bulk electrolyte outside the  nanometer sized equilibrium double layer is electroneutral. Increasing the voltage results in a diffusion-limited value of current. A non-electroneutral space charge layer starts to form outside the double layer at higher voltages. On further increasing the voltage beyond a critical value, the ion-concentration and potential profiles are no longer one-dimensional and electrolyte flow is observed due to a hydrodynamic instability known as electroconvection. The current rises rapidly beyond the diffusion limited value, resulting in over-limiting current.  According to a model developed by \citet{rubinstein2001electro}, tangential electric fields interact with the space charge and this drives electroconvection.  Experimental results obtained by \citet{rubinstein2008direct} show the 3 different regimes in a current-voltage plot - the Ohmic regime, limiting current and non-equilibrium over-limiting current due to electroconvection.
Rubinstein and Zaltzman \cite{rubinstein2001electro,rubinstein2000electro,rubinstein2005electroconvective,lerman2005absence,zaltzman2007electro,rubinstein2007electro}  used a slip boundary condition and linear stability analysis to predict the critical voltage for the onset of electroconvection. It has been shown that instability of the bulk electrolyte and instability due to electro-osmotic slip of the first kind cannot lead to electroconvection \cite{lerman2005absence,zholkovskij1996electrokinetic}. Electro-osmotic slip of the first kind occurs due to the  action of tangential electric fields on charges in the double layer. According to the second kind electro-osmotic slip model developed by  \citet{rubinstein2001electro}, it is the action of tangential electric forces on the space charge that results in a slip velocity at the outer edge of the space charge layer. This slip velocity drives electroconvection in the bulk electrolyte. Linear stability analysis with the second kind slip velocity leads to a shortwave catastrophe as the critical voltage for the onset of electroconvection decreases monotonically and asymptotically to a shortwave limiting value.  Solving the full electroconvective problem resolves this issue \cite{rubinstein2005electroconvective,rubinstein2007electro}.  Linear stability analysis using the slip boundary condition has also been used to solve problems where the effect of polymer additives and surface kinetics on electroconvection have been investigated \cite{tikekar2018electroconvection,li2022electroconvection}.  \citet{tikekar2018electroconvection} used linear stability analysis along with the slip boundary condition \cite{rubinstein2001electro} to analyze a system with high molecular weight polymers added to the electrolyte. They showed that the addition of entangled polymers to the electrolyte results in an increase in the critical voltage required for the onset of electroconvection.  \citet{li2021suppression,li2022electroconvection} used the ultraspherical spectral method \cite{olver2013fast} to solve the linear stability analysis problem in the entire domain  instead of using the slip boundary condition. The ultraspherical spectral method \cite{olver2013fast} was needed to fully resolve the ion concentration, potential and velocity profiles in the thin double layer and the space charge layer. \citet{li2021suppression} showed that an externally imposed cross    	flow can increase the critical voltage for the onset of electroconvection. \citet{li2022electroconvection} studied interfacial kinetics and showed that slower reaction rates at the ion-selective surface can increase the critical voltage required for the onset of instability.

Previously, various techniques have been used to suppress electroconvection, such as the use of solid electrolytes, use of polymer additives or the modification of the solid-electrolyte interface (SEI). Porous alumina separators with electrolyte in the pores can stop electroconvection, which results in a more uniform deposition \cite{tu2014nanoporous,tu2015dendrite,lu2014stable,kang2014improved}. Solid electrolytes such as ceramics suffer from problems such as poor ionic conductivity and high interfacial resistance due to irregular contact between electrolyte and electrode. Such electrolytes might also be brittle and fragile \cite{zheng2020regulating}.  \citet{wei2018stabilizing} added high molecular weight polymers to the electrolyte to increase viscosity, which suppresses electroconvection. They showed that adding $10\%$ PMMA to the electrolyte increases the critical voltage for the onset of electrocovnection to 1V, and the conductivity is not significantly affected. To further increase the critical voltage, a higher volume fraction of polymer would be required, which would result in a decrease in the conductivity of the electrolyte. Modification of the SEI can be done in two ways - polymer additives can get adsorbed on the surface of the electrode, or the electrode surface is modified before it is used. For example, \citet{markevich2017fluoroethylene} used fluoroethylene carbonate (FEC) based electrolytes to modify the SEI, while \citet{tu2017designing} added ionomers to the electrode surface to improve the performance of lithium batteries. However, all of these techniques involve significant modifications of the bulk electrolyte. The technique proposed in this study involves the addition of low molecular weight polymers to the electrolyte,  which does not change the properties of the bulk electrolye significantly. Van der Waals forces of attraction between the metal electrode and the dissolved polymers results in the formation of a thin layer of higher concentration near the electrode. The polymers in this layer remain dissolved in solution. Transport of ions through the electrolyte remains unhindered.  


In this study, we use linear stability analysis to analyze the effect of the dissolved polymer layer on electroconvection. The van der Waals potential is modeled by using the Hamaker constant \cite{hamaker1937london,israelachvili2011intermolecular} and considering the interaction between a particle and a flat surface. Previously, van der Waals forces have been modeled using a Hamaker constant when studying the spread of surfactant monolayers over thin films \cite{matar1999spreading,jensen1992insoluble}. This method has also been used to describe rupture of liquid films \cite{ruckenstein2018spontaneous} and interaction between  particles in bubbling fluidized bed reactors \cite{shrestha2020effect}.   The van der Waals potential results in the formation of a thin layer of dissolved  polymers near the electrode surface. In section \ref{sec:2}, we solve for the polymer concentration in the base state using an approach similar to that used by \citet{russel1991colloidal} to describe concentration profiles in sedimentation. A soft sphere model \cite{cohen2009phenomenological,li2015equation} is used to calculate the osmotic pressure as a function of the volume fraction of the polymer. In section \ref{sec:3}, we analyze the linear stability of the system and show that the dissolved polymer layer opposes the growth of perturbations.

\section{\label{sec:2}PROBLEM SETUP}

\subsection{Solution domain}
\begin{figure}[htbp]
	\includegraphics[scale=.5]{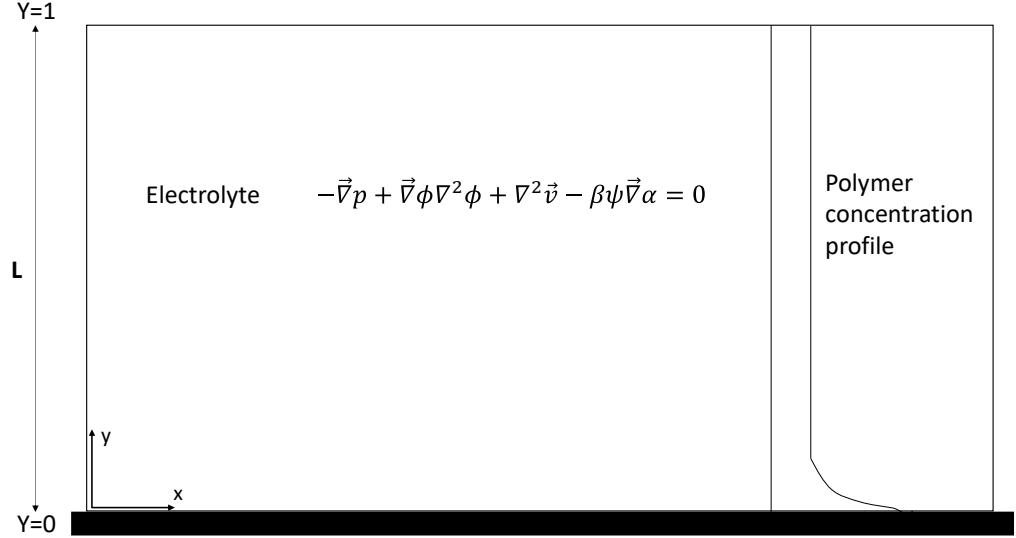}
	\caption{ Schematic of the system. The electrolyte is enclosed between an ion-selective surface (y = 0) and a stationary reservoir (y = 1), with an adsorbed polymer layer at y = 0.}
	\label{fig:sdomain}
    \end{figure}

The electrolyte is bounded by a metal electrode at $y=0$ and a stationary reservoir at $y=1$. The system is shown in figure \ref{fig:sdomain}. The electrode at the opposite end is assumed to be far enough that the electrolyte near it is not affected by electroconvection, hence it is replaced by a stationary reservoir.    The electrolyte is an aqueous solution of a binary univalent electrolyte. Cations move towards the metal electrode when the  battery is being charged.  Low molecular weight polymers are dissolved in the electrolyte. The polymers move towards the metal electrode due to van der Waals forces of attraction and form a thin layer close to the ion-selective surface. The polymers in this layer remain dissolved in solution and play an important role in suppressing  electroconvection, which is driven by the action of tangential electric fields on space charges adjacent to the electrode. Van der Waals forces on the dissolved polymers results in a body force in the space charge region that opposes the flow of the electrolyte. 

\subsection{Governing equations}
The governing equations are the Nerst–Planck equations for ion transport, the Poisson equation for electric potential, the Stokes equation for electrolyte flow and species transport equation for the dissolved polymer. $C^+$ and $C^-$ are the concentrations of the cation and anion and $\phi$ is the electric potential, $\mathbf{u}$ is the velocity, $p$ is the pressure and $\psi$ is the polymer concentration.  $D=\frac{D^+}{D^-}=1$ is the ratio of the diffusivities of the cation and anion in the electrolyte. $\delta=\frac{\sqrt[]{\frac{\epsilon_r \epsilon_0RT}{2F^2C_0}}}{L}$ is the non-dimensional double layer thickness, where $\epsilon_0$ is the vacuum permittivity, $\epsilon_r$ is the dielectric constant, $R$ is the universal gas constant, $T$ is the absolute temperature, $F$ is Faraday's constant, $C_0$ is the initial ion concentration in the electrolyte and $L$ is the inter-electrode distance. $C_0$ and $L$ are used to non-dimensionalize ion concentration and lengths. $D_0 = \frac{2D^+D^-}{D^++D^-}$ is the average salt diffusivity. $U = \frac{\epsilon_r\epsilon_0(RT)^2}{F^2\eta L}$ is the velocity scale and is used to non-dimensionalize velocity. $\eta$ is the fluid viscosity. The P\'eclet number is defined as $Pe = \frac{UL}{D_0}$. Pressure and time are non-dimensionalized using the scales  $p_0 = \frac{\eta U}{L}$ and  $t_0=\frac{L^2}{D_0}$. The thermal voltage $\frac{RT}{F}$ is used to non-dimensionalize electric potential.  Similar definitions for the non-dimensional numbers have been used by  \citet{rubinstein2005electroconvective} and \citet{li2022electroconvection}. The P\'eclet number is set to 0.5, which is a typical value for aqueous salt solutions. $\delta=0.001$ is the non-dimensional double layer thickness. This value for the double layer thickness ensures that the double layer and the space charge layer are much smaller than the distance between the ion-selective surface and the reservoir. This ensures that the system can accurately predict the physics of the problem while keeping the computational cost within reasonable limits. $\delta=0.001$ lies within the ranges calculated by \citet{rubinstein2005electroconvective} and \citet{li2022electroconvection}.  $\alpha$ is the potential due to van der Waals forces of attraction between the polymer molecules and the metal electrode. $\psi$ is the polymer concentration. The van der Waals potential is non-dimensionalized using the factor $k_BT$ where $k_B$ is the Boltzmann constant. The polymer concentration is non-dimensionalized using the initial concentration $\psi_0$, which is also the concentration of polymer in the stationary reservoir at $y=1$.

The  non-dimensional equations are:

\begin{subequations}
	\label{eqn:governing}
	\begin{equation}
		\frac{\partial C^+}{\partial t} + Pe\mathbf{u}.\boldsymbol{\nabla} C^+ = \frac{D+1}{2}\boldsymbol{\nabla}.(\boldsymbol{\nabla} C^+ + C^+\boldsymbol{\nabla} \phi)
	\end{equation}
	\begin{equation}
		\frac{\partial C^-}{\partial t} + Pe\mathbf{u}.\boldsymbol{\nabla} C^- = \frac{D+1}{2D}\boldsymbol{\nabla}.(\boldsymbol{\nabla} C^- - C^-\boldsymbol{\nabla} \phi)
	\end{equation}
	\begin{equation}
		-2 \delta^2\nabla^2 \phi = c^+ - c^-
	\end{equation}
	\begin{equation}
		\boldsymbol{\nabla}.\mathbf{u}=0
	\end{equation} 
	\begin{equation}
		-\boldsymbol{\nabla} p+ \nabla^2 \mathbf{u} + \nabla^2\phi \boldsymbol{\nabla} \phi -\beta \psi \boldsymbol{\nabla}\alpha  = 0
	\end{equation}
	\begin{equation}
	\frac{\partial \psi}{\partial t} + Pe\mathbf{u}.\boldsymbol{\nabla} \psi = D_P\boldsymbol{\nabla}.(d_p\left(\phi_P\right)\boldsymbol{\nabla}\psi + \psi\boldsymbol{\nabla} \alpha)
	\end{equation} 
    \label{eqn:GE}
\end{subequations}
Equation \ref{eqn:GE}e is obtained by adding the force due to van der Waals potential to the Stokes equation. The Reynold's number in electroconvection tends to 0 so the inertial terms have been neglected. The Schmidt number ($=\frac{\eta}{\rho_0 D_0}$) is generally large, which means that the velocity is effectively quasi-steady as it changes on a time scale smaller than ion-concentrations. $\beta$ is a non-dimensional term. $\beta=\frac{\psi_0 \alpha_{sc} L}{\eta U}=\frac{k_BT L}{\eta U}=5\times10^9$. $\alpha_{sc}=k_BT$ and $\psi_0=1 mole/m^3$ are scalings for the potential and polymer concentration. $k_B$ is the Boltzmann constant.  In equation \ref{eqn:GE}f, $D_P=0.1$ is the diffusivity of the polymer relative to that of ions,  when the polymer is dilute. $d_p\left(\phi_P\right)$ adds the effect of variation in the volume fraction $\phi_P$ of the polymer on its diffusivity, which is important near the electrode where the polymer concentration is higher. $\phi_P$ is the volume fraction of a sphere encircling a polymer with  radius equal to the radius of gyration $R_g$ of the polymer. $d_p\left(\phi_P\right)$ is expressed as:
\begin{equation}
d_p\left(\phi_P\right)=\frac{d}{d\phi_P}\left[\phi_PZ\left(\phi_P\right)\right]
\end{equation}
$Z=\frac{\pi}{nkT}$ is the compressibility factor. $\pi$ is the osmotic pressure. An expression for the osmotic pressure is obtained from \citet{cohen2009phenomenological} and  \citet{li2015equation}. They derived an expression for the osmotic pressure of polyethylene glycol (PEG) which is valid in both the dilute and semi-dilute regimes. The following expression is used for the osmotic pressure:
\begin{equation}
	\pi N^{9/5}=\frac{RT}{M_m\bar{V}}\left[\left(\frac{C}{C^*_N}\right)+0.49\left(\frac{C}{C^*_N}\right)^{9/4}\right]
\end{equation}
$C$ is the mass concentration of the polymer. $C^*_N \approx N^{-4/5}/\bar{V}$ is the overlap concentration of a polymer with $N$ monomers. It is the concentration at which the polymer transitions from the dilute to semi-dilute regime. $\bar{V}$ is the partial specific volume of the polymer. Each polymer molecule occupies a sphere with radius equal to the radius of gyration $R_g$ of the polymer. $M_m$ is the molar mass of the monomer. The above equation is used to obtain the expression for the compressibility factor:
\begin{equation}
     Z=\frac{\pi}{nk_BT}=1+0.5\phi_P^{5/4}
\end{equation}
The volume fraction of the polymer is expressed as the ratio of the concentration to the overlap concentration $\phi_P=\frac{C}{C^*_N}$. As the polymer concentration is higher than the overlap concentration in the semi-dilute regime, the volume fraction of polymer spheres ($\phi_P$) will be greater than 1 in this case due to overlap.

\subsection{Expression for van der Waals potential}
\begin{figure}[htbp]

	\centering
	\includegraphics[width=1\textwidth]{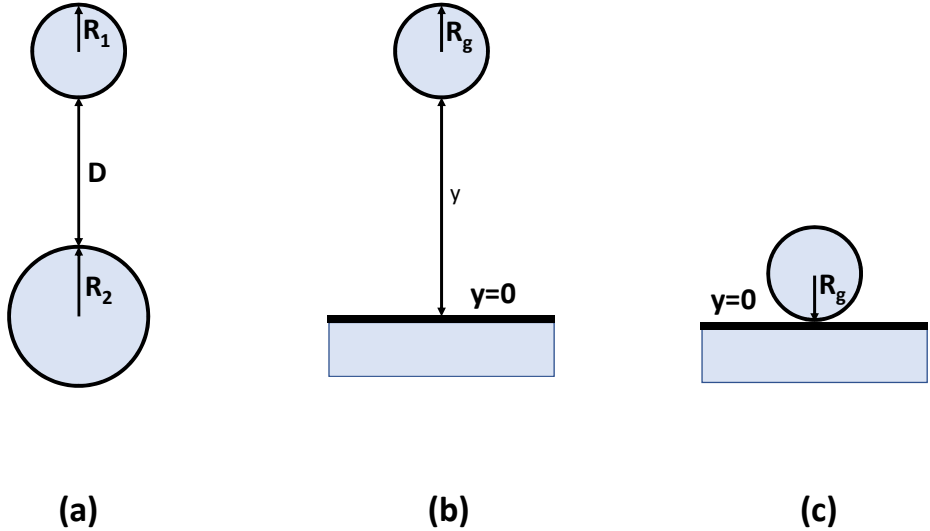}
	
	\caption{Calculation of Hamaker constant. (a) General case with two spheres separated by a distance D (b) Since the metal electrode has a flat surface, $R_2\to\infty$  (c) When $y=0$, the denominator is 0, so $R_g$ is added to the denominator as a correction, in order to avoid an unbounded potential at $y=0$.}
	\label{fig:A}
\end{figure}

\begin{figure}[htbp]
	\begin{minipage}[htbp]{0.45\linewidth}
		
		\centering
		\includegraphics[width=\textwidth]{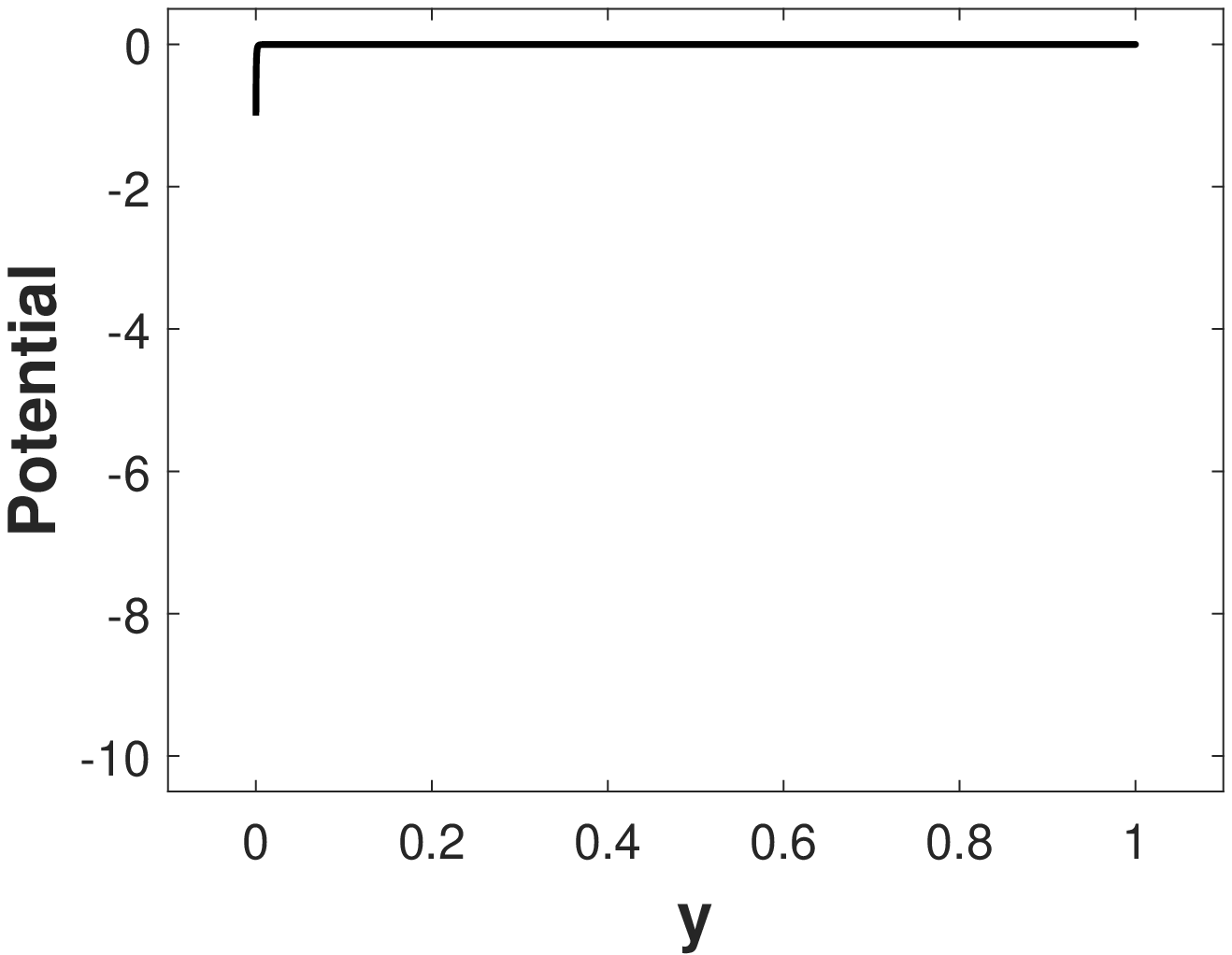}
		(a)
	\end{minipage}
	\hspace{0.5cm}
	\begin{minipage}[htbp]{0.45\linewidth}
		\centering
		\includegraphics[width=\textwidth]{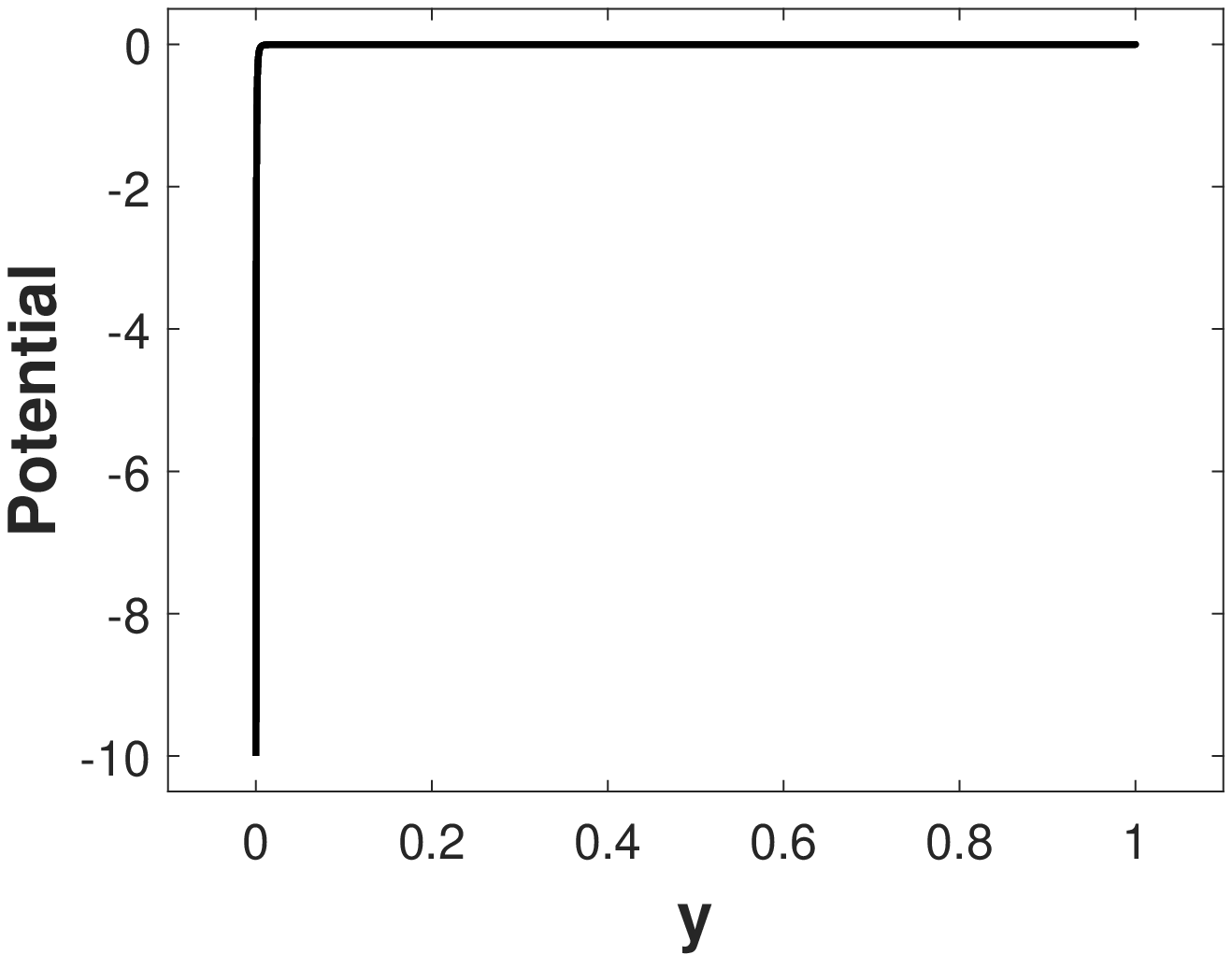}
		(b)
	\end{minipage}
	\caption{van der Waals potential. In these figures, the radius of gyration $R_g$ is 0.001. The Hamaker constant A is (a) 1 (b) 10. }
	\label{fig:base_Vpot}
\end{figure}

To obtain an expression for the van der Waals potential responsible for the attraction between the polymers and the metal electrode, we start with the expression for the force of attraction between two spheres of radius $R_1$ and $R_2$ with their surfaces separated by a distance $D$. In such a scenario, the van der Waals potential is defined as \cite{israelachvili2011intermolecular}:
\begin{equation}
	\alpha=-\frac{A'}{6}\left[\frac{2R_1R_2}{(2R_1+2R_2+D)D}+\frac{2R_1R_2}{(2R_1+D)(2R_2+D)}+ln\left(\frac{D(2R_1+2R_2+D)}{(2R_1+D)(2R_2+D)}\right)\right]
\end{equation} 
$A'$ is the Hamaker constant. Since the metal electrode has a flat surface, its radius of curvature is infinity ($R_2\to\infty$). As mentioned before, the polymer molecules are assumed to occupy a sphere with radius equal to the radius of gyration of the polymer ($R_1=R_g$). It is also assumed that the distance between the polymer and the metal electrode is much larger than the radius of the sphere encircling the polymer molecule $\left(\frac{R_1}{D}<<1\right)$. The distance $D$ is the sum of the normal coordinate $y$ and the radius of gyration $R_g$. The radius of gyration is added as a correction to prevent the denominator from going to 0 at the electrode surface and to set $\alpha=-A$ at $y=0$. This prevents the potential from growing to infinity at $y=0$, so that the problem can be solved numerically. The potential profile still has the correct variation with distance from the electrode, so the physics of this problem can be studied using this model. At minimum separation, $y=0$ and the distance between the center of the sphere and the electrode surface is $R_g$.  Figure \ref{fig:A} is  a schematic showing the spheres and the distances mentioned above. Using the above equations and assumptions, the expression for the van der Waals potential can be simplified to:

\begin{equation}
	\alpha=-\frac{2A'}{9}\left(\frac{R_1}{D}\right)^3=-A\left(\frac{R_g}{(y+R_g)}\right)^3
	\label{eqn:VdPot}
\end{equation} 
$A=\frac{2A'}{9}$ is a modified constant. The non-dimensional Hamaker constant $A'$ can be expressed as $A'=\frac{\pi^2Cn_1n_2}{k_BT}$. Here $n_1$ and $n_2$ are the number densities of polymer molecules and metal atoms. C is the coefficient of the attraction term in Lenard-Jones potential $-C/r^6$. $n_2=8.5\times10^{28} \frac{molecules}{m^3}$. This is obtained using the density of copper ($8.96 \frac{g}{cm^3}$) and the molar mass of copper ($63.55\frac{g}{mol}$). $n_1=6.023\times10^{23}$ is calculated by using a reference concentration for the polymer - $1 \frac{mole}{m^3}$. Higher polymer concentrations are reflected by the polymer concentration $\psi$ in the body force term $\beta\psi\boldsymbol{\nabla}\alpha$ in the Stokes equation.  $C=4\epsilon_{Cu-PEG}\sigma_{Cu-PEG}^6$, where $\epsilon$ is a measure of interaction strength and $\sigma$ is a measure of interaction distance. The values of $\epsilon_{Cu-PEG}$ and $\sigma_{Cu-PEG}$ are approximated as $\epsilon_{Cu-PEG}=\sqrt{\epsilon_{Cu-Cu}\epsilon_{PEG-PEG}}$ and $\sigma_{Cu-PEG}=\sqrt{\sigma_{Cu-Cu}\sigma_{PEG-PEG}}$ \cite{sebeck2016alkane,grimme2006semiempirical}. The values of bond energy and bond distance are obtained from \citet{sebeck2016alkane} for copper and \citet{lee2009coarse} for PEG/PEO (poly-ethylene oxide). Substituting these values. we get a value of A equal to 2. Since we considered only one example of metal and polymer in this calculation, the actual value of the Hamaker constant can vary. In our calculations, we consider two values for the Hamaker constant: $A=1$ and $A=10$, representing interactions weaker and stronger than the case for which we calculated A.  The potential due to van der Waals forces is calculated (equation \ref{eqn:VdPot}) and shown in figure \ref{fig:base_Vpot}. The value of the radius of gyration  is considered to be 0.001(dimensional value - $10^{-6}m$), which is larger than the usual values for polymers. This is done to ensure that the numerics remain tractable inside the double layer. Both the double layer and the radius of gyration are slightly larger than typical values, however both are significantly smaller than the inter-electrode distance, hence they are small enough to describe the physics accurately. 
\subsection{Conversion of polymer concentration to volume fraction}
The polymer concentration and volume fraction are related as:
\begin{equation}
	\psi=\frac{\phi_P}{\frac{4}{3}\pi R_g^3N_A}
\end{equation}
$R_g$ is the radius of gyration and $N_A$ is the Avogadro's number. The value of $R_g$ is taken to be 1 nm in this calculation only, to reflect more realistic values for the conversion factor between concentration and volume fraction. Substituting, we get $\psi = \frac{\phi_P}{0.0025} \frac{moles}{m^3}$ 

	\subsection{\label{subsec:23} Boundary conditions}
The boundary conditions are:
\begin{subequations}
	\begin{equation}
		C^+|_0=C_s, C^-|_0=0, C^+|_1=C_s\textquotesingle, C^-|_1=C_s\textquotesingle
	\end{equation}
	\begin{equation}
		\phi|_0=0, \phi|_1=V_0
	\end{equation}
	\begin{equation}
		\mathbf{u}|_0=0, \mathbf{u}|_1=0
	\end{equation}
	\begin{equation}
		\mathbf{n}.(\boldsymbol{\nabla}C^- - C^- \boldsymbol{\nabla}\phi )|_0=0
	\end{equation}
	\begin{equation}
 D_P(d_p\left(\phi_P\right)\boldsymbol{\nabla}\psi + \psi\boldsymbol{\nabla} \alpha)|_0 =0, \psi|_1=\psi_0
    \end{equation}
	\label{eqn:full_BC}
\end{subequations}

 We use equilibrium boundary conditions at $y = 0$ ($C^+|_0=C_s, C^-|_0=0$). $C_s$ is the cation concentration at $y = 0$. At $y=1$, the ion concentrations are fixed $\left(=C_s\textquotesingle\right)$.  $C_s=1$ and $C_s\textquotesingle=1$.  The anion flux is 0 at the electrode $\left(y=0\right)$ and at the reservoir $\left(y=1\right)$ since the total amount of anions in the electrolyte remains constant. The electrode and reservoir are maintained at potentials 0 and $V_0$. The velocity is 0 at $y=0$ (no-slip boundary condition). The electrolyte in the reservoir is at rest, so the velocity at $y=1$ is also 0. Periodic boundary conditions  are used in the direction tangential to the electrode surface $\left(x=0,6\right)$. Similar governing equations and boundary conditions have been used in previous studies \cite{rubinstein2005electroconvective,li2021suppression,li2022electroconvection}. The metal electrode does not allow polymer to pass through it so the polymer flux is 0 at $y=0$. The amount of polymer at $y=1$ remains fixed and is equal to the initial polymer concentration $\left(=\psi_0\right)$.

\section{\label{sec:3}Linear stability analysis}

In this section, we use linear stability analysis to calculate the growth rates of perturbations. We solve the governing equations in the entire domain including the thin double layer and space charge layer adjacent to the ion-selective surface. \citet{rubinstein2005electroconvective}, \citet{rubinstein2007electro}, \citet{li2021suppression} and \citet{li2022electroconvection} have used linear stability analysis to predict the critical voltage for the onset of electroconvection. In these studies, the thin double layer was resolved in order to obtain accurate values for the critical voltage. Since ion concentrations and velocity have to be solved in these thin layers where the variables vary rapidly, accurate eigenvalue solvers are required. We use the ultraspherical spectral method \cite{olver2013fast}, which has also been used to solve problems involving electroconvection \cite{li2021suppression,li2022electroconvection}.  The ultraspherical method leads to matrices that are almost banded and can be used to solve linear ordinary  differential equations with variable coefficients. The eigenvalue problem is solved to obtain the growth rates for the onset of instability as a function of wavenumber of perturbations.

\subsection{Base state}
In the base state, the variables are a function of the normal coordinate only and there is no velocity. Substituting $\mathbf{u}=0$ and considering variations in the normal coordinate $y$ only, the governing equations (equation \ref{eqn:GE}) can be used to obtain the base state equations and the boundary conditions.

\noindent \textbf{Governing equations in the electrolyte ($y=0-1$):}

 \begin{subequations}
\begin{equation}
	\frac{dC^+}{dy} + C^+ \frac{d\Phi}{dy} = I
\end{equation}
\begin{equation}
	\frac{dC^-}{dy} - C^-\frac{d\Phi}{dy} = 0
\end{equation}
\begin{equation}
	-2\delta^2 \frac{d^2 \Phi }{dy^2} = C^+ - C^-
\end{equation}
\begin{equation}
	d_p\left(\phi_P\right)\frac{d\Psi}{dy} + \Psi\frac{d\alpha}{dy}=0
\end{equation}
\label{eqn:BASE_full}
\end{subequations}   

\noindent \textbf{Boundary conditions at  $y=0$:}

\begin{equation}
	\left.C^+\right\rvert_0=1, \left.C^-\right\rvert_0=0, \left.\Phi\right\rvert_0=0,  \left. \left(\frac{\partial C^-}{\partial y} - C^- \frac{\partial \Phi}{\partial y}\right)\right\rvert_0=0
\end{equation}

\begin{figure}[htbp]
	\begin{minipage}[htbp]{0.45\linewidth}
		
		\centering
		\includegraphics[width=\textwidth]{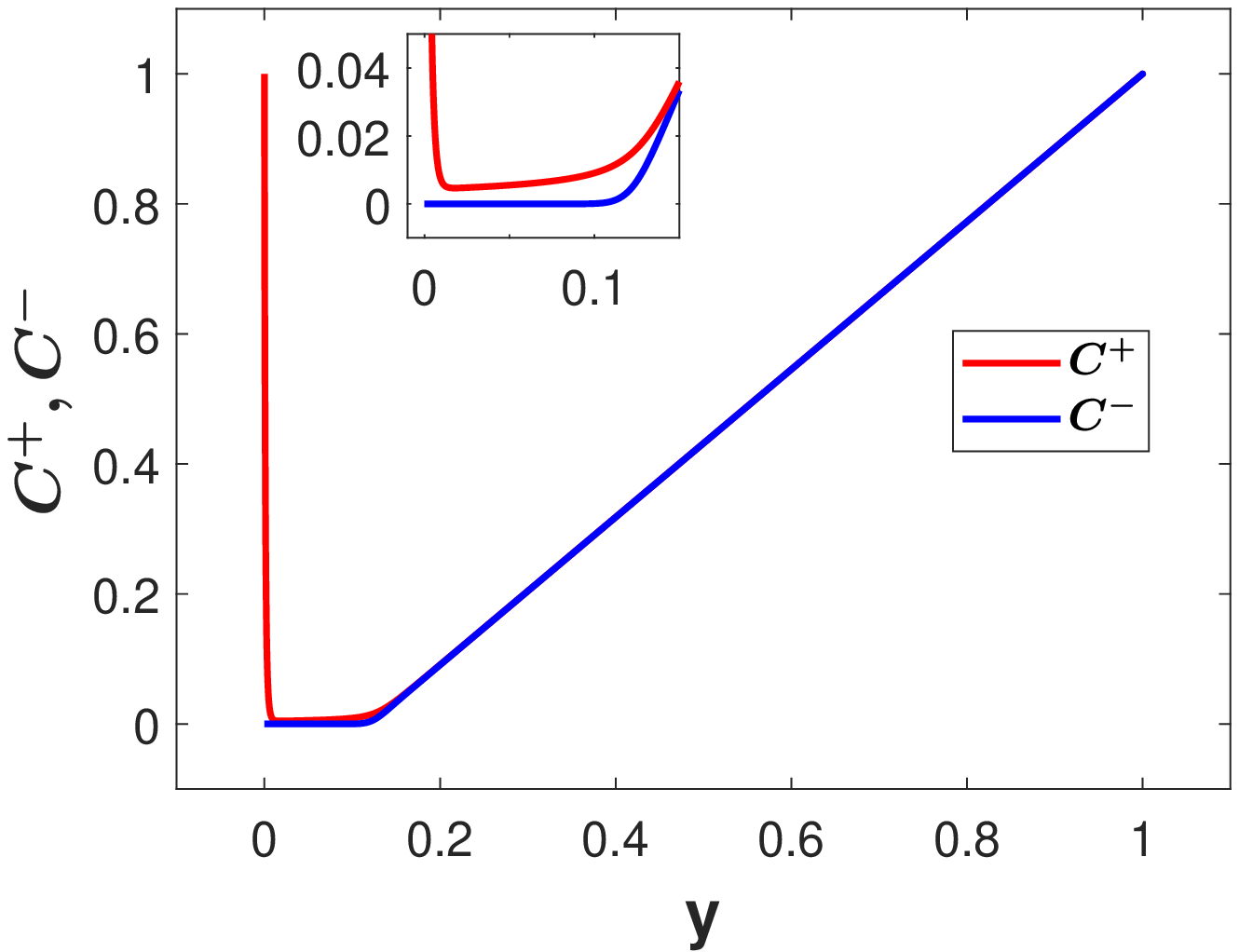}
		(a)
	\end{minipage}
	\hspace{0.5cm}
	\begin{minipage}[htbp]{0.45\linewidth}
		\centering
		\includegraphics[width=\textwidth]{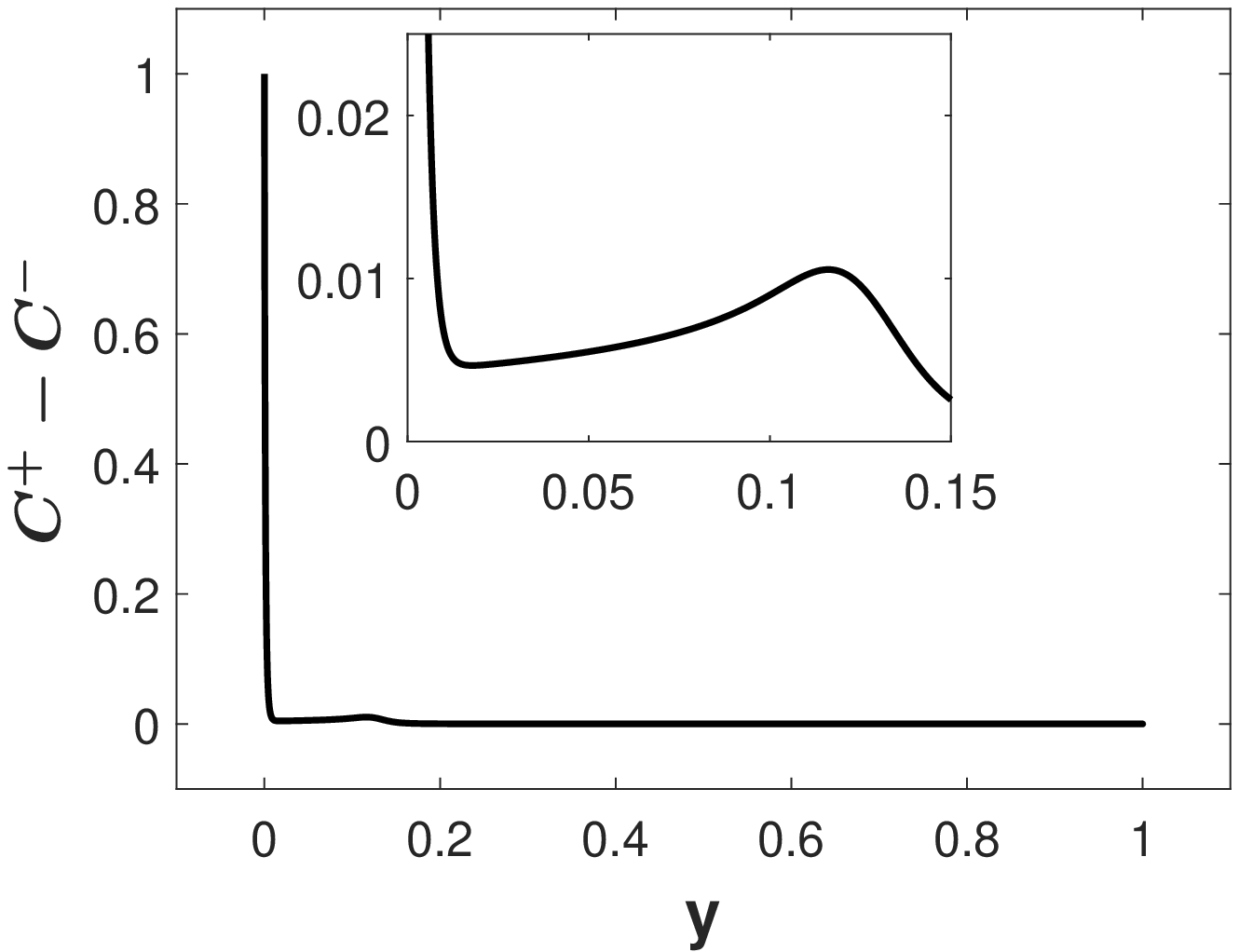}
		(b)
	\end{minipage}
	\caption{(a) Ion concentrations and (b) Net charge ($C^+-C^-$) in the base state at V = 50 and $\delta=0.001$. The non-dimensional current is 1.1364. The insets show a magnified view of the plots in the space charge layer, between the double layer and the bulk electrolyte. The space charge layer thickness ($\delta_s$) is considered to be the distance from the electrode to the peak in the concentration profile. In this case, $\delta_s=0.116$. }
	\label{fig:base_state}
\end{figure}

\begin{figure}[htbp]
	\begin{minipage}[htbp]{0.45\linewidth}
		
		\centering
		\includegraphics[width=\textwidth]{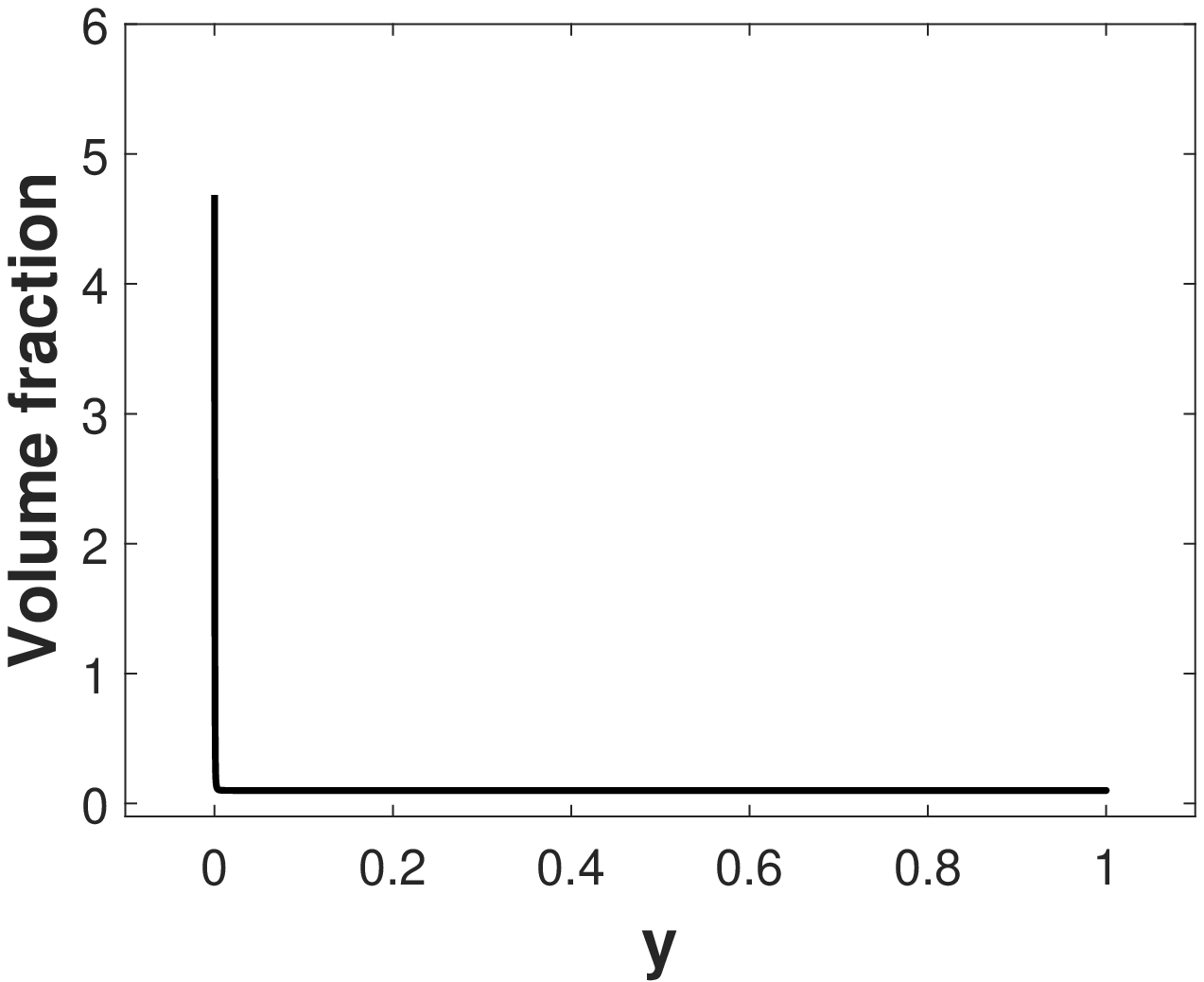}
		(a)
	\end{minipage}
	\hspace{0.5cm}
	\begin{minipage}[htbp]{0.45\linewidth}
		\centering
		\includegraphics[width=\textwidth]{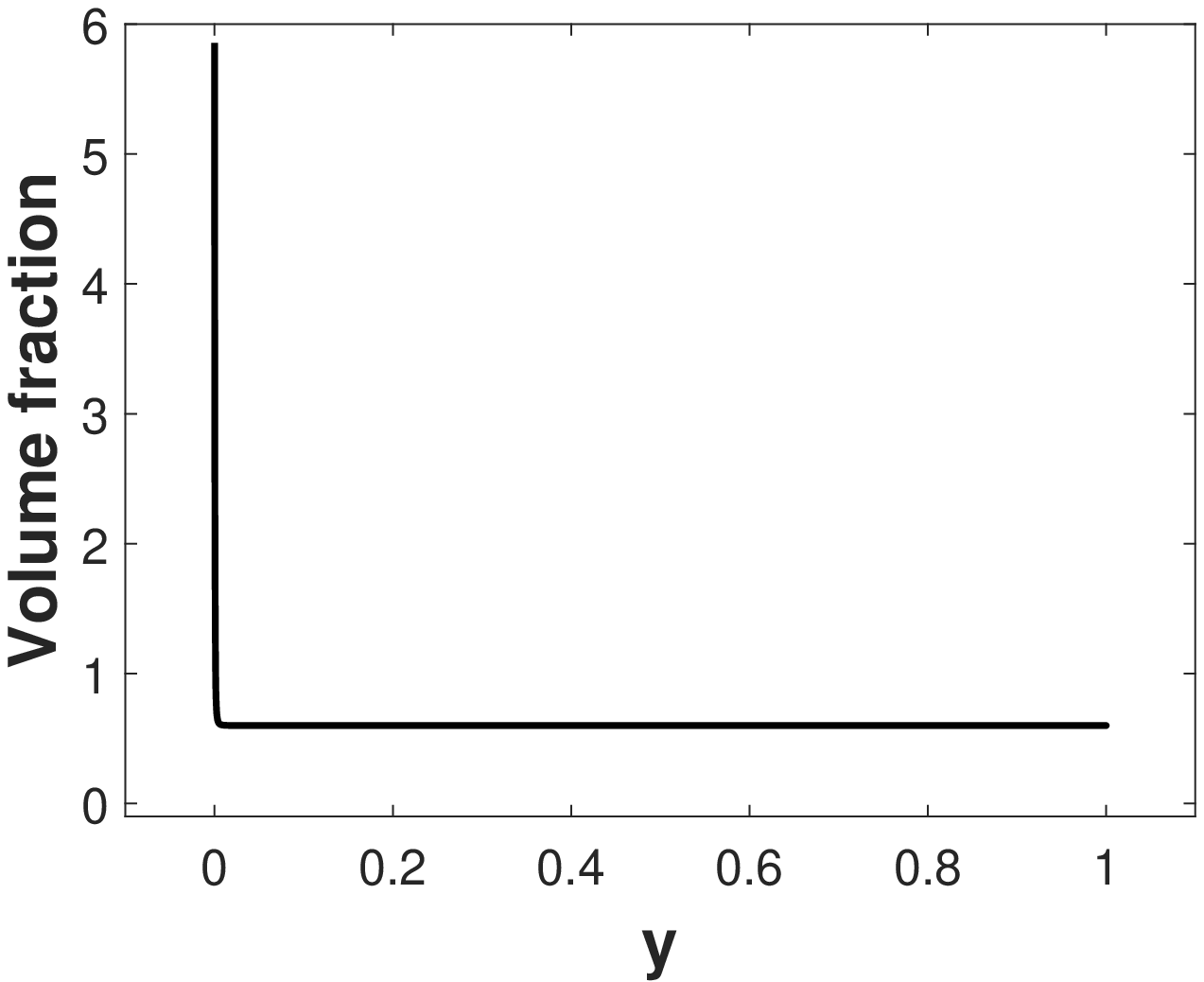}
		(b)
	\end{minipage}
	\caption{Polymer volume fractions in the base state. In these figures, the Hamaker constant A is 10, the radius of gyration $R_g$ is 0.001 and the bulk volume fraction of the polymer is (a) 0.1 (b) 0.6 . }
	\label{fig:base_pol}
\end{figure}

 \noindent \textbf{Boundary conditions at  $y=1$:}
 
\begin{equation}
	\left.C^+\right\rvert_1=1,  \left.C^-\right\rvert_1=0, \left.\Phi\right\rvert_1=V_0,\Psi\rvert_1=\psi_0
\end{equation}
The ion concentration profile in the base state is shown in figure \ref{fig:base_state}. The polymer concentration profile is shown in figure \ref{fig:base_pol}.

\subsection{Perturbed equations}

We perturb the variables and calculate the growth rate of the perturbations. The perturbations are defined as follows:
	\begin{equation}
	\left[
	\begin{array}{l}
		c\left(x,y'',t\right) \\
		\phi\left(x,y'',t\right) \\
		p\left(x,y'',t\right) \\
		u\left(x,y'',t\right) \\
		v\left(x,y'',t\right)\\
		\psi\left(x,y'',t\right)
	\end{array}
	\right]
	=
	\left[
	\begin{array}{l}
		\hat{c}\left(y''\right) \\
		\hat{\phi}\left(y''\right) \\
		\hat{p}\left(y''\right) \\
		\hat{u}\left(y''\right) \\
		\hat{v}\left(y''\right)\\
		\hat{\psi}\left(y''\right)
	\end{array}
	\right]
	\exp\left(\sigma t +ikx\right)
	\label{eqn:perturbed_quantities}
\end{equation}

Substituting the perturbations in the governing equations (equation \ref{eqn:GE}) and the boundary conditions , we obtain the   following  equations:

\noindent \textbf{Governing equations in the electrolyte ($y=0-1$):}
\begin{subequations}
	\begin{equation}
		\sigma \hat{c}^++Pe\hat{v}\frac{dC^+}{dy}=\frac{D+1}{2}\left(\frac{d^2\hat{c}^+}{dy^2}-k^2\hat{c}^+-k^2C^+\hat{\phi}+\frac{d}{dy}\left(C^+\frac{d\hat{\phi}}{dy}\right)+\frac{d}{dy}\left(\hat{c}^+\frac{d\Phi}{dy}\right)\right)
	\end{equation}
	\begin{equation}
		\sigma \hat{c}^-+Pe\hat{v}\frac{dC^-}{dy}=\frac{D+1}{2D}\left(\frac{d^2\hat{c}^-}{dy^2}-k^2\hat{c}^-+k^2C^-\hat{\phi}-\frac{d}{dy}\left(C^-\frac{d\hat{\phi}}{dy}\right)-\frac{d}{dy}\left(\hat{c}^-\frac{d\Phi}{dy}\right)\right)
	\end{equation}
	\begin{equation}
		2\delta^2 \left(\frac{d^2 \hat{\phi}}{d y^2}-k^2\hat{\phi}\right) = \hat{c}^- - \hat{c}^+
	\end{equation}
	\begin{equation}
		\frac{d^4 \hat{v}}{dy^4}-2k^2\frac{d^2 \hat{v}}{dy^2}+k^4\hat{v}=k^2\left(\left(\frac{d^2 \hat{\phi}}{d y^2}-k^2\hat{\phi}\right)\frac{d\Phi}{dy}-\hat{\phi}\frac{d^3\Phi}{dy^3}-\beta\hat{\psi}\frac{d\alpha}{dy}\right)
	\end{equation}
\begin{equation}
	\hat{v}\left(-Pe\frac{d\Psi}{dy}\right)+\hat{\psi}\left(D_p\frac{d^2\alpha}{dy^2}-k^2D_pd_p\left(\phi_P\right)\right)+\frac{d\hat{\psi}}{dy}\left(D_p\frac{d\alpha}{dy}+D_p\frac{dd_P(\phi_P)}{dy}\right)+\frac{d^2\hat{\psi}}{dy^2}\left(D_pd_P(\phi_P)\right)=\sigma\hat{\psi}
\end{equation}
	\label{eqn:P_full_l1}
\end{subequations}   
\noindent \textbf{Boundary conditions at  $y=0$:}
\begin{subequations}
	\begin{equation}
		\left.\hat{c}^+\right\rvert_0=0
	\end{equation}
	\begin{equation}
		\left. \left(\frac{d \hat{c}^-}{dy} - C^- \frac{d\hat{\phi}}{dy}- \hat{c}^- \frac{d\Phi}{dy}\right)\right\rvert_0=0
	\end{equation}
	\begin{equation}
		\left. \hat{\phi} \right\rvert_0 = 0
	\end{equation}
	\begin{equation}
		\left.	\hat{v} \right\rvert_0=0 , \left.	\frac{d\hat{v}}{dy} \right\rvert_0=0
	\end{equation}
\begin{equation}
\frac{d\alpha}{dy}\left. \hat{\psi}\right\rvert_0+\left.d_P(\phi_P)\frac{d \hat{\psi}}{dy}\right\rvert_0=0
\end{equation}
	\label{eqn:full_y_0}
\end{subequations}
\noindent \textbf{Boundary conditions at  $y=1$:}
\begin{subequations}
	\begin{equation}
		\left.\hat{c}^+\right\rvert_1=0
	\end{equation}
	\begin{equation}
		\left.\hat{c}^-\right\rvert_1=0
	\end{equation}
	\begin{equation}
		\left. \hat{\phi} \right\rvert_1 = 0
	\end{equation}
	\begin{equation}
		\left.	\hat{v} \right\rvert_1=0 , \left.	\frac{d\hat{v}}{dy} \right\rvert_1=0
	\end{equation}
	\begin{equation}
		\hat{\psi}\rvert_1=0
	\end{equation}
\end{subequations}
The above equations are solved using the ultraspherical spectral method \cite{olver2013fast} to obtain the growth rates, which are the eigenvalues of this set of equations.

\subsection{Results}

    \begin{figure}[htbp]

	\centering
	\includegraphics[width=0.5\textwidth]{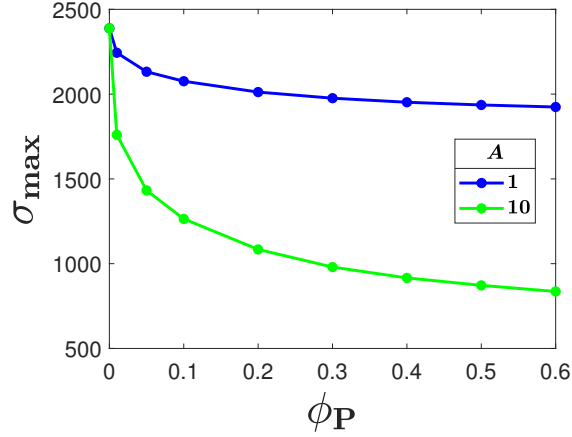}
	
	\caption{Maximum growth rate at different values of Hamaker constant, as a function of the bulk volume fraction. In these figures, $V=50$ and $\delta=0.001$. }
	\label{fig:BlkVF}
\end{figure}

        	\begin{figure}[htbp]
	\begin{minipage}[htbp]{0.5\linewidth}
		\centering
		\includegraphics[width=\textwidth]{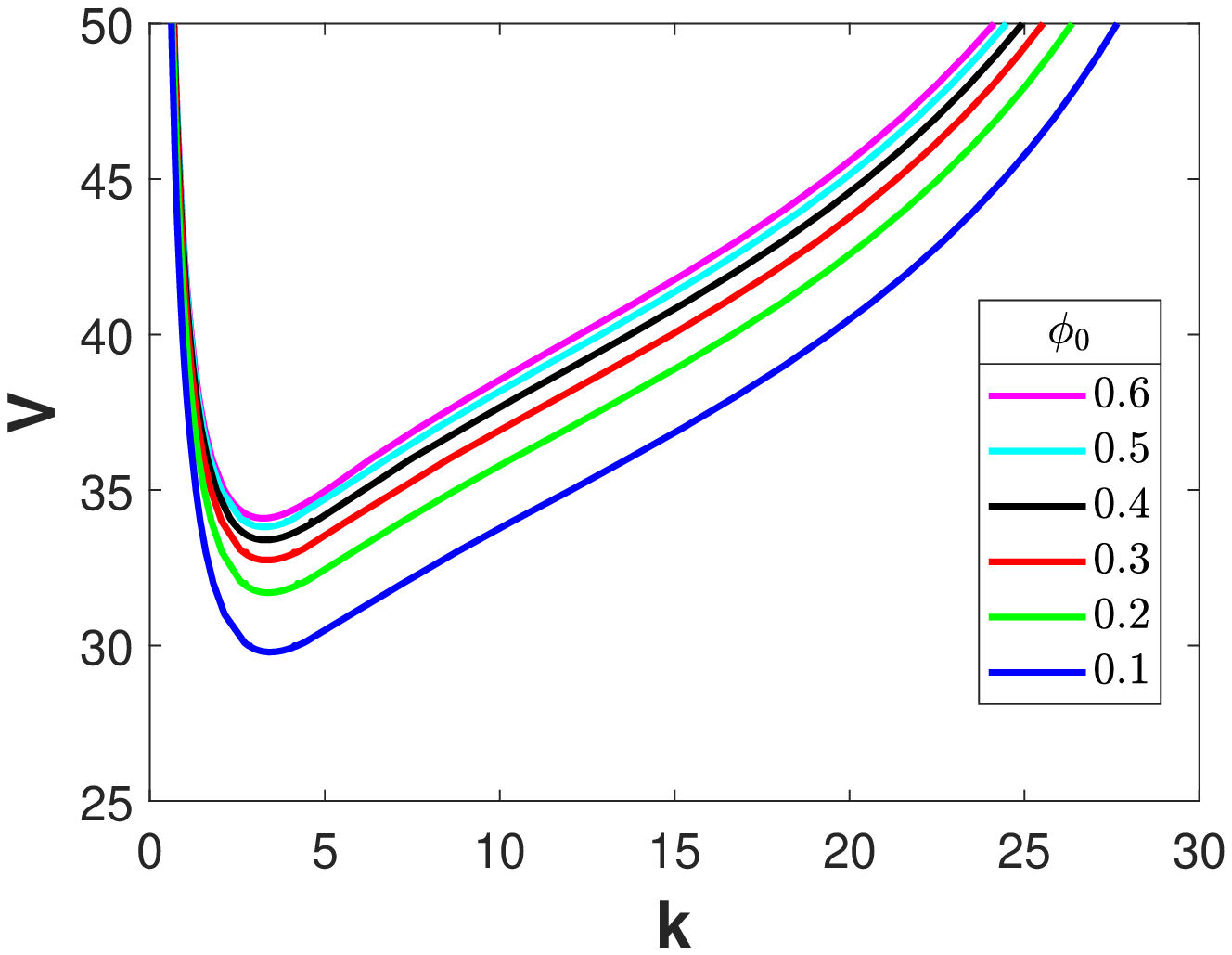}
		(a)
	\end{minipage}
	\hspace{0.2cm}
	\begin{minipage}[htbp]{0.47\linewidth}
		
		\centering
		\includegraphics[width=\textwidth]{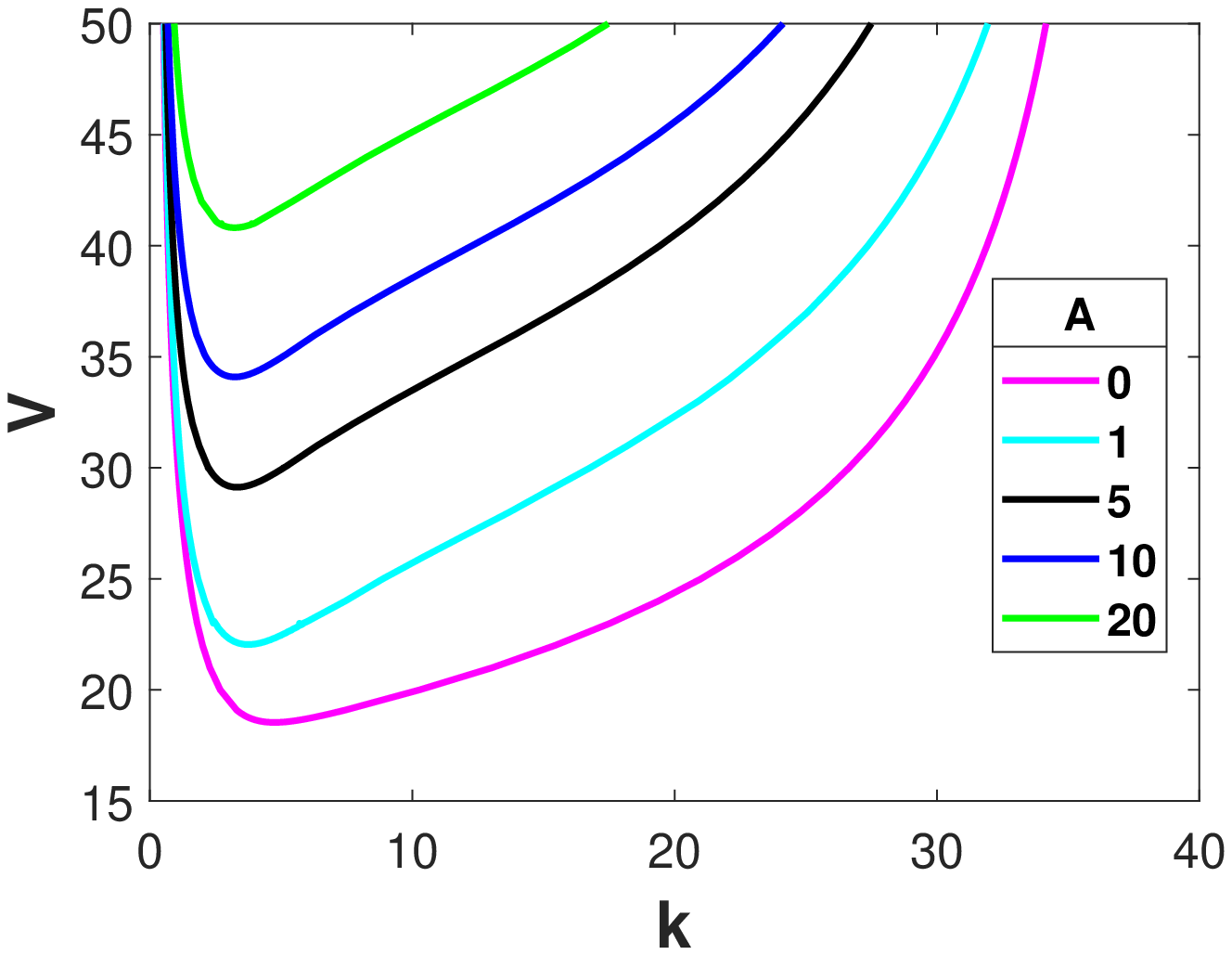}
		(b)
	\end{minipage}
	\hspace{0.2cm}
	
	\caption{Neutral stability plots with varying (a) bulk volume fraction (Hamaker constant = 10) (b) Hamaker constant (bulk volume fraction = 0.6). In these figures, $V=50$ and  $\delta=0.001$.  }
	\label{fig:MS}
\end{figure}
In figure \ref{fig:BlkVF}, the maximum growth rate of perturbations is plotted as a function of the bulk volume fraction, which is also the initial volume fraction of the polymer. The amount of polymer dissolved in the electrolyte near the ion-selective surface increases with an increase in the bulk volume fraction or an an increase in the strength of the van der Waals interactions (represented by an increase in the Hamaker constant). An increase in the amount of polymer near the electrode increases the stabilizing effect of the dissolved polymer, resulting in a decrease in the maximum growth rate as seen in figure \ref{fig:BlkVF}. This also results in an increase in the critical voltage required for the onset of electroconvection, as seen in figure \ref{fig:MS}. In figure \ref{fig:MS}, the critical voltage for the onset of instability is plotted as a function of wavenumber, for different values of the bulk volume fraction (figure \ref{fig:MS}a) and for different values of the Hamaker constant (figure \ref{fig:MS}b). 
    \begin{figure}[htbp]

	\centering
	\includegraphics[width=\textwidth]{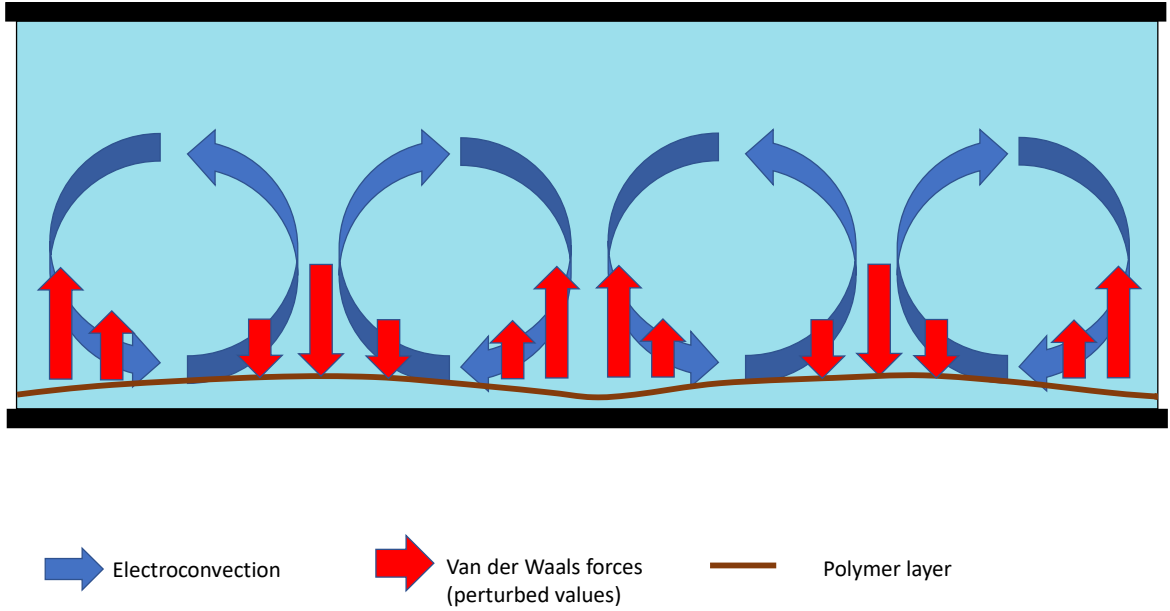}
	
	\caption{Mechanism responsible for the suppression of electroconvection. }
	\label{fig:mechanism}
\end{figure}

Figure \ref{fig:mechanism} shows a schematic explaining the stabilization of the system due to the dissolved polymer layer. The blue arrows show the electroconvective vortices. The brown line shows an isosurface of polymer concentration. It can be seen that where the electrolyte moves upwards, the polymer concentration is higher than the base state. This means that the perturbed polymer concentration $\hat{\psi}$ is positive, and the body force $-\beta\hat{\psi}\boldsymbol{\nabla}\alpha$
is negative. This means that when the electrolyte moves upwards, the body force due to van der Waals attraction points downwards and opposes the upwards motion of the electrolyte. Similarly, when the electrolyte flows downwards, $\hat{\psi}$ is negative due to a lower polymer concentration, resulting in an upward force that opposes the downwards motion of the electrolyte. The body force due to van der Waals potential acts a restoring force that opposes electroconvective flow, and hence, opposes the growth of perturbations.

\section{Conclusion}

Adding low molecular weight polymers can increase the critical voltage required for the onset of electroconvection without changing the properties of the bulk electrolyte significantly. This is because of van der Waals forces of attraction between the metal electrode and the dissolved polymers, which results in the formation of  a thin layer of higher polymer concentration near the ion-selective surface.  The polymers in this layer remain dissolved in solution. The van der Waals force on this dissolved polymer layer results in a restoring body force that opposes the growth of perturbations, making the system more stable and increasing the critical voltage required for the onset of electroconvection.

	\bibliographystyle{unsrtnat}	
\bibliography{vdw}

\end{document}